\documentclass[aip,pof,graphicx,10pt]{revtex4-1}
\usepackage{amssymb,amsbsy,times,fancyhdr,color}
\usepackage{amsmath}
\usepackage{mathrsfs,amsfonts}
\usepackage{latexsym,afterpage}
\usepackage{graphicx,subfigure,multirow}
\usepackage{color}
\usepackage{color}
\usepackage{amsmath}
\newcommand{\dy}{\partial}
\newcommand\ddy[2]{\frac{\dy#1}{\dy#2}}

\newcommand{\ex}{\mathrm{e}}

\newcommand{\zi}{{\rm i}}

\newcommand{\grad}{\nabla}

\newcommand{\eb}{\boldsymbol{e}}

\newcommand{\ub}{\boldsymbol{u}}

\newcommand{\Ubar}{{\overline{U}}}

\newcommand{\Qbar}{{\overline{Q}}}
\newcommand{\rhobar}{{\overline{\rho}}}

\newcommand{\qhat}{{\hat{q}}}

\newcommand{\Ri}{\widehat{R}}
\newcommand{\Fr}{F}

\definecolor{dark-green}{rgb}{0,0.5,0} % leeds

\begin{document}

%\preprint{}

\title{Stratified shear flow instabilities in the non-Boussinesq regime} 


%\thanks{}

\author{E. Heifetz}

\author{J. Mak}
\email[]{julian.c.l.mak@googlemail.com}
\altaffiliation{School of Mathematics, The University of Edinburgh, James
Clerk Maxwell Building, The King's Buildings, Edinburgh, EH9 3FD, UK}

\affiliation{Department of Geosciences, Tel Aviv University, Tel Aviv, 69978,
Israel}

%\date{\today}

\begin{abstract}

Effects of the baroclinic torque on wave propagation normally neglected under
the Boussinesq approximation is investigated here, with a special focus on the
associated consequences for the mechanistic interpretation of shear instability
arising from the interaction between a pair of vorticity-propagating waves. To
illustrate and elucidate the physical effects that modify wave propagation, we
consider three examples of increasing complexity: wave propagation supported by
a uniform background flow; wave propagation supported on a piecewise-linear
basic state possessing one jump; and an instability problem of a
piecewise-linear basic state possessing two jumps, which supports the
possibility of shear instability. We find that the non-Boussinesq effects
introduces a preference for the direction of wave propagation that depends on
the sign of the shear in the region where waves are supported. This in turn
affects phase-locking of waves that is crucial for the mechanistic
interpretation for shear instability, and is seen here to have an inherent
tendency for stabilisation.

\end{abstract}


\maketitle %\maketitle must follow title, authors, abstract and \pacs


%===============================================================================

\section{Introduction}\label{s1}

An approximation that one often makes when studying the dynamics of stratified
fluids is the Boussinesq approximation\cite{Salmon-GFD, Vallis-GFD}. One assumes
that the variation of density about a background reference density
$\rho'/\rho_0$ is small, and thus we may neglect inertial effects associated
with such terms except when it is multiplied by the gravitational acceleration
$g$, i.e., buoyancy effects dominate. This assumption of small density deviation
is well satisfied in the ocean and remains useful for studying certain
atmospheric flows. The Boussinesq equation has and still remains a useful model
for investigating a variety of fluid dynamical phenomena in geophysical systems,
such as convection\cite{Chandrasekhar-Stability, LohseXia10}, wave-mean flow
interaction\cite{Buhler-Waves} and shear
instabilities\cite{DrazinReid-Stability, Carpenter-et-al13}, the last of which
will be our principal focus here.

To study shear instabilities, one often makes a further simplifying assumption
by employing `defects' in the velocity and/or density profile (i.e.,
piecewise-constant/linear profiles) as a model for sharp gradients in the basic
state. Such an assumption is useful for the study of the onset of instabilities
for several reasons: the resulting dispersion relation often reduces to a low
order algebraic equation, for which analytical as well as asymptotic solutions
exist; such solutions are often the leading asymptotic solution for general
smooth profiles in the long-wave limit\cite{DrazinHoward62,
DrazinReid-Stability}; there is a mechanistic interpretation for the
instability, seen as the constructive interference of vorticity propagating
waves travelling counter to the background flow\cite{Carpenter-et-al13}. The use
of defects has life beyond linear theory, allowing the derivation of reduced
models via matched asymptotic methods to investigate the nonlinear development
and saturation of shear instabilities\cite{Balmforth-et-al97,
Balmforth-et-al12}. Since the use of defects is as a model for sharp gradients
in the basic state, one can ask whether it might be more appropriate to study
flow instabilities in the presence of sharp density gradients without the
Boussinesq approximation, since the assumption of small density variation may no
longer hold. To this end, there have been several works studying shear
instabilities beyond the Boussinesq approximation over the years, using smooth
profiles but with a density that has a small scale height\cite{MasloweKelly71},
and classic profiles with defects in\cite{UmurhanHeifetz07, BarrosChoi11,
BarrosChoi14}. However, these aforementioned works in the non-Boussinesq setting
focuses on solving the modified Taylor--Goldstein equation to investigate the
property of growth rates with increasing deviation from the Boussinesq regime
(which will be seen to be measured by a Froude number), without necessarily
providing a physical reason of what causes the modifications to the instability
characteristics. Our work here aims to complement these previous works by
investigating the mechanistic modifications to the underlying wave dynamics by
non-Boussinesq effects, and how this affects the mechanistic interpretation of
the instability accordingly. We provide mathematical details and physical
schematics on how the part of the baroclinic torque neglected by the Boussinesq
approximation generates vorticity anomalies; how this affects wave propagation
and interaction is illustrated for increasingly more complex examples. An
instability problem where the cause of instability is strongly affected by the
non-Boussinesq term is then presented and analysed accordingly.

The layout of the document is as follows. In Section~\ref{s2} we formulate the
problem in terms of the vorticity, displacement and pressure, to relate the
generation of vorticity anomalies by the Boussinesq and non-Boussinesq effects.
The dynamics of waves supported on a uniform background is investigated and
rationalised in Section~\ref{s3}, to illustrate some of the possible effects due
to the non-Boussinesq term. In Section~\ref{s4}, we consider a more complex
example where waves are now supported on defects, and rationalise also the
changes induced by the non-Boussinesq term. In Section~\ref{s5}, a slightly
simpler version of the Taylor--Caulfield instability\cite{Taylor31, Caulfield94,
Carpenter-et-al10a, Rabinovich-et-al11, Carpenter-et-al13} in the non-Boussinesq
regime is investigated and analysed accordingly. This ties together the
modification to the wave dynamics and the instability properties resulting from
the action-at-a-distance interaction between non-Buossinesq interfacial waves.
We conclude and discuss our results in Section~\ref{s6}.

%==============================================================================

\section{Mathematical formulation}\label{s2}

\begin{figure}[tb]
\begin{center}
	\includegraphics[width=\textwidth]{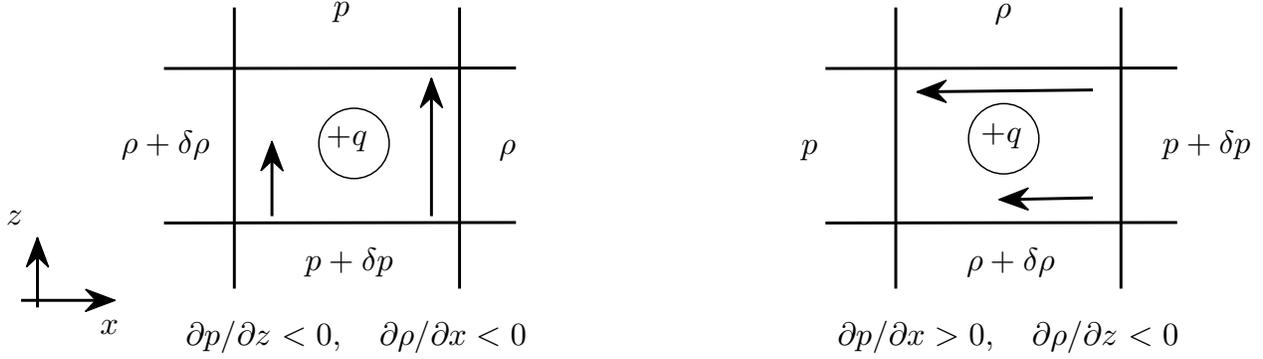}
	\caption{Vorticity generation via baroclinic torque associated with $p$ and
	$\rho$ anomalies. We show here a case where $(\dy p/\dy z)(\dy\rho/\dy x)>0$,
	a case resembling hydrostatic balance, and a case where $-(\dy p/\dy
	x)(\dy\rho/\dy z)>0$. Pressure gradients results in a tendency in flow
	acceleration flow. However, the acceleration is inversely proportional to the
	density of the fluid parcel, thus it results in shear and hence vorticity
	anomalies.}
	\label{fig:prho_vorticity}
\end{center}
\end{figure}

We assume a two-dimensional, inviscid incompressible flow in the ($x,z$) plane,
with governing equations
\begin{equation}
	\frac{D\ub}{Dt}=-g\eb_{z}-\frac{1}{\rho}\grad p,\qquad
	\frac{D\rho}{Dt}=0,
\end{equation}
where $\ub=(u,0,w)$ is the velocity, $g$ is the gravitational acceleration, $p$
is the kinematic pressure, $\rho$ is the density, and $D/Dt = \dy/\dy t +
\ub\cdot\grad$ is the material derivative. The last equation for density comes
from assuming $\grad\cdot\ub=0$. Defining $q=\dy w/\dy x-\dy u/\dy z$ (note that
the vorticity component in the $y$ direction is $\omega_{y}=-q$), the $q$
equation is given by
\begin{equation}
	\frac{Dq}{Dt}=\eb_y\cdot\left(\frac{1}{\rho^{2}}\grad p\times\grad\rho\right)=
	\frac{1}{\rho^{2}}\left(\ddy{p}{z}\ddy{\rho}{x}
	-\ddy{p}{x}\ddy{\rho}{z}\right),
\end{equation}
How both terms of the baroclinic torque generate vorticity anomalies is
illustrated in the schematic depicted in Figure~\ref{fig:prho_vorticity}, with
the details in the caption.

Suppose we take a basic state $\Ubar(z)$, with $\Qbar(z)=-\dy\Ubar/\dy z$. Then
it follows from the $x$-component of the momentum equation that we may take
$\overline{p}=\overline{p}(z)$. This leads to the basic state satisfying the
hydrostatic balance
\begin{equation}\label{s2:hydrostatic}
	\ddy{\overline{p}}{z}=-\rhobar g,
\end{equation}
and we consider a basic state $\rhobar=\rhobar(z)$. With this, we observe that
the terms in the square brackets above will be $O(\epsilon)$ once we linearise,
and so the contributions at $O(\epsilon)$ from $\rho^{-2}$ is $(\rhobar)^{-2}$.
A linearisation of the vorticity equation then results in
\begin{equation}
	\left(\ddy{}{t}+\Ubar\ddy{}{x}\right)q=
	-w\ddy{\Qbar}{z}-\frac{1}{\rhobar^{2}}
	\left(\rhobar g\ddy{\rho}{x}+\ddy{\rhobar}{z}\ddy{p}{x}\right),
\end{equation}
where the quantities with no overbars are perturbation quantities. Linearising
the continuity equation results in the system of equations given by
\begin{equation}\label{s2:system1}
	\left(\ddy{}{t}+\Ubar\ddy{}{x}\right)q
	=-w\ddy{\Qbar}{z}-\frac{g}{\rhobar}\ddy{\rho}{x}
	-\frac{1}{\rhobar^{2}}\ddy{\rhobar}{z}\ddy{p}{x},\qquad
	\left(\ddy{}{t}+\Ubar\ddy{}{x}\right)\rho
	=-w\ddy{\rhobar}{z},
\end{equation}
upon using the divergence-free condition on the perturbation velocity field.
Note that the last term on the right hand side of the linearised $q$ equation is
absent in the Boussinesq limit.

With the vertical perturbation displacement defined as
\begin{equation}
	\left(\ddy{}{t}+\Ubar\ddy{}{x}\right)\zeta=w,
\end{equation}
an integration yields the identity
\begin{equation}\label{s2:rhozeta}
	\rho=-\ddy{\rhobar}{z}\zeta,
\end{equation}
where only advective effects are considered. Since the velocity field is assumed
to be non-divergent, we may define a streamfunction such that
\begin{equation}\label{s2:stream1}
	u=-\frac{\dy\psi}{\dy z},\qquad w=\frac{\dy\psi}{\dy x},
\end{equation}
and this results in the identity $\grad^{2}\psi=q$. Defining the Buoyancy
frequency to be $N^{2}=-(g/\rhobar)(\dy\rhobar/\dy z)$, the system of equations
\eqref{s2:system1} becomes
\begin{equation}\label{s2:system2}
	\left(\ddy{}{t}+\Ubar\ddy{}{x}\right)q
	=-\ddy{\psi}{x}\ddy{\Qbar}{z}
	-N^{2}\ddy{}{x}\left(\zeta-\frac{1}{g}\frac{p}{\rhobar}\right), \qquad
	\left(\ddy{}{t}+\Ubar\ddy{}{x}\right)\zeta=\ddy{\psi}{x}.
\end{equation}
The $(\dy/\dy x)(p/\rhobar)$ term is the correction that is absent in the
Boussinesq regime. Here, $\psi$ may be formally inverted from $q$ via a Green's
function, which depends on the chosen domain and boundary conditions, so in
theory we have a formulation in terms of $q$ and $\psi$, once we substitute for
$p$ in some way. The pressure $p$ will be seen to be related to $\psi$ and thus
$q$ via a substitution from the momentum equation.

We observe that there are three
dynamical regimes:
\begin{enumerate}
  \item A barotropic regime where $\zeta= p/(g\rhobar)$. For barotropic flow,
  $p=p(\rho)$, so that
  \begin{equation}
    \frac{Dp(\rho)}{Dt} = \frac{\mathrm{d}p}{\mathrm{d}\rho}\frac{D\rho}{Dt}.
  \end{equation}
  With incompressibility, $D\rho/Dt=0$, and so $p=-\zeta(\dy\overline{p}/\dy
  z)$. Upon using the hydrostatic balance relation \eqref{s2:hydrostatic}, we
  recover the identity $\zeta= p/(g\rhobar)$. In this regime, the linearised
  baroclinic torque cancels out exactly, consistent with the assumption of
  barotropicity.

	\item The Boussinesq regime where $\zeta\gg p/(g\rhobar)$. In this regime the
	fluid parcel adjusts its pressure distribution to the surrounding environment
	on a fast enough time-scale such that pressure effects may be neglected, and
	buoyancy effects dominate.
	
	\item The case where $\zeta\ll p/(g\rhobar)$. In this case, the non-Boussinesq
	effects outweigh the buoyancy effects and become the dominant player in the
	dynamics.
\end{enumerate}
We will now consider related examples of increasing complexity to see how the
extra non-Boussinesq term in the linearised baroclinic torque influences the
dynamics. 

%==============================================================================

\section{Basic wave dynamics}\label{s3}

We consider first the case where waves are supported on a uniform background
flow with $N=\textnormal{constant}$. Without loss of generality, we take
$\Ubar=0$, and the governing equations \eqref{s2:system2} reduces to
\begin{equation}\label{s3:system1}
	\ddy{q}{t}=-N^2\ddy{}{x}\left(\zeta - \frac{1}{g}\frac{p}{\rhobar}\right),
	\qquad \ddy{\zeta}{t}=\ddy{\psi}{x}.
\end{equation}
To substitute for $p$, we turn to the momentum equation $\dy u/\dy
t=-(1/\rhobar)(\dy p/\dy x)$. Substituting for $p$ in equation
\eqref{s3:system1}, taking another time-derivative of the vorticity equation and
substituting for $\dy\zeta/\dy t$ results in
\begin{equation}\label{s3:equ1}
	\ddy{^2}{t^2}\left(\grad^2\psi-\frac{N^2}{g}\ddy{\psi}{z}\right)
	= -N^2\ddy{^2\psi}{x^2}.
\end{equation}
This is the Taylor--Goldstein equation for this simplified case.

Now, taking $\rhobar = \rho_0 \ex^{-z/H}$, $\dy\rhobar/\dy z=-\rhobar/H$, so
$N^2 = -(g/\rhobar)(\dy\rhobar/\dy z)=g/H$, where $H$ is a density scale height.
Equation \eqref{s3:equ1} becomes
\begin{equation}
	\ddy{^2}{t^2}\left(\grad^2-\frac{1}{H}\ddy{}{z}\right)\psi
	= -N^2\ddy{^2\psi}{x^2}.
\end{equation}
With modal solutions of the form $\psi=\hat{\psi}(z)\ex^{\zi k(x-ct)}$, we
obtain the dispersion relation
\begin{equation}\label{s3:equ2}
	c^2\left(\ddy{^2}{z^2}-\frac{1}{H}\ddy{}{z}-k^2\right)\hat{\psi} = -N^2\psi.
\end{equation}
We see that solutions of the form $\hat{\psi}\sim\ex^{\zi mz}\ex^{z/(2H)}$
satisfies $|\ub|^2\sim\ex^{z/H}$ and $\rhobar\sim\ex^{-z/H}$, so
$\rhobar|\ub|^2<\infty$. Substituting this form of solution into equation
\eqref{s3:equ2}, the imaginary parts cancel out exactly, and the resulting
dispersion relation is given by
\begin{equation}\label{s3:dispersion}
	c^2=\frac{N^2}{m^2 + k^2 + 1/(4H^2)}.
\end{equation}
In the Boussinesq limit, $H\to\infty$, and we recover the usual dispersion
relation for gravity waves in a non-rotating system\cite{Vallis-GFD}. For
$H<\infty$, the phase speed of the waves reduced via the $1/(4H^2)$ term. This
is akin to how the scale height affects acoustic-gravity waves\cite{YehLiu74},
and similar to the way in which the existence of a finite Rossby deformation
radius attenuates the phase speed of Rossby waves\cite{HeifetzCaballero14}.

\begin{figure}[tb]
\begin{center}
	\includegraphics[width=0.9\textwidth]{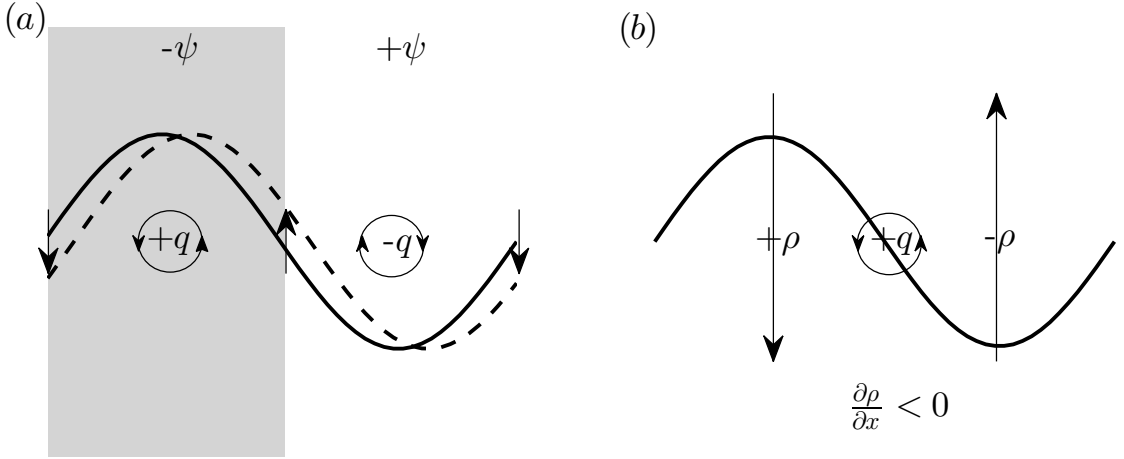}
	\caption{Schematic for gravity waves propagation in the Boussinesq limit.
	($a$) Direction of propagation depends on how $\zeta$ and $q$ are correlated
	at the peaks and troughs. Shown here is the case where $\zeta\sim q$, and
	results in $c>0$; with $\zeta\sim -q$, the opposite is true. ($b$) Vorticity
	anomalies at the nodes resulting from the baroclinic torque associated with
	the Boussinesq term, which depends on the $\zeta$ configuration and not on the
	direction of wave propagation.}
	\label{fig:bous_wave}
\end{center}
\end{figure}

A physical reason for this reduced phase speed may be rationalised via the
changes to vorticity anomalies generated by the corresponding baroclinic
torques. For simplicity, we consider the case with $m=0$, and, for completeness,
we consider the Boussinesq limit first. The vorticity generation comes from the
$(\dy\overline{p}/ \dy z)(\dy\rho/\dy x)=-g\rhobar(\dy\rho/\dy x)$ term; since
$g\rhobar>0$, the sign of the resulting vorticity anomalies is correlated with
the sign of $-(\dy\rho/\dy x)$. The direction of the wave propagation is
dependent on how $\zeta$ is correlated with $q$; $\zeta\sim q$ gives rightward
propagating waves with $c>0$, as in Figure~\ref{fig:bous_wave}($a$), and vice
versa\cite{Harnik-et-al08}. However, regardless of direction of wave
propagation, equation \eqref{s2:rhozeta} indicates that, for stable
stratification, the $\zeta$ distribution sketched in
Figure~\ref{fig:bous_wave}($b$) results in $\dy\rho/\dy x<0$, and thus this
results in positive vorticity anomaly at the node of the wave. Another way of
thinking about it is that the peaks of the wave has the tendency to descend
whilst the troughs wants to rise, therefore the resulting movement in this case
is anti-clockwise, and is thus a positive vorticity anomaly as in
Figure~\ref{fig:prho_vorticity}($a$).

\begin{figure}[tb]
\begin{center}
	\includegraphics[width=0.9\textwidth]{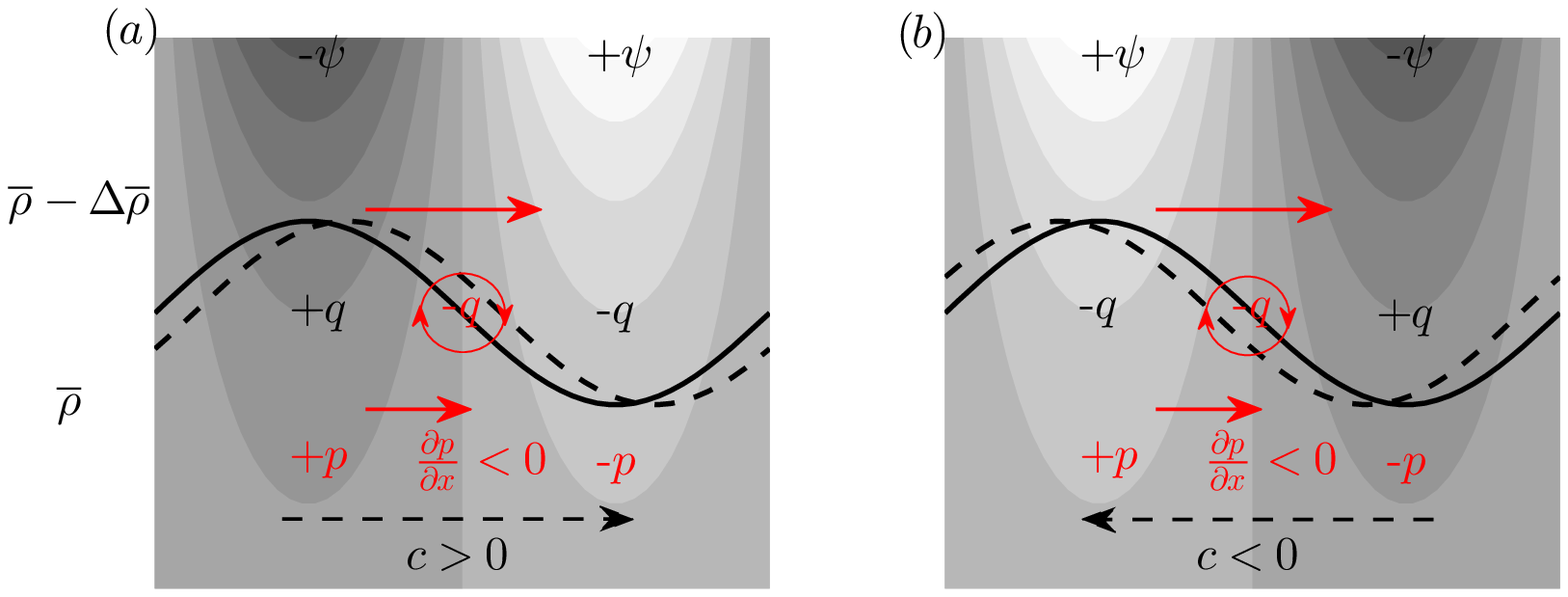}
	\caption{Schematic of the torque generated by the non-Boussinesq contribution
	for gravity waves with $m=0$. The imposed wave structure is as in
	Figure~\ref{fig:bous_wave}, the left and right panels depict a right and
	leftward propagation wave respectively, and the shading is associated with
	contours of $\psi=\textnormal{const}$. As a result of the vorticity
	distribution at the peaks and troughs, we have the corresponding pressure
	distributions which in turn generates vorticity anomalies at the wave nodes;
	these are seen to be the opposite sign as the one associated with the
	Boussnesq term shown in Figure~\ref{fig:bous_wave}($b$).}
	\label{fig:non_bous_wave}
\end{center}
\end{figure}

In the more general case with the non-Boussinesq term we also need to work out
the distribution of $p$ and see how this modifies the scenario depicted in
Figure~\ref{fig:bous_wave}. From the $x$-momentum equation, for a right-going
wave ($c>0$), observing that $\psi\sim\ex^{z/(2H)}$, we have
\begin{equation}\label{s3:pq}
  -cu=c\ddy{\psi}{z}=-\frac{p}{\rhobar}\qquad\Rightarrow\qquad
  p\sim-\ddy{\psi}{z}\sim-\psi\sim q,
\end{equation}
and thus $p\sim q$\footnote{For the purpose of this study, we assume that the
scale height is large enough so that $k>1/(2H)$. The attenuation of the wave
propagation for the anomalous case where $k<1/(2H)$ and $\psi\sim q$ can also be
rationalised using similar arguments, but will not be discussed here.}. This
scenario is depicted in Figure~\ref{fig:non_bous_wave}($a$). The resulting $p$
distribution leads to $\dy p/\dy x<0$ at the nodes, and thus the vorticity
anomaly generated is related to $-(\dy\rhobar/\dy z)(\dy p/\dy x)<0$, resulting
in a negative vorticity anomaly at the node as in
Figure~\ref{fig:prho_vorticity}($b$). So the correction torque results in
vorticity anomalies that is of the opposite sign to the one generated by
Boussinesq term depicted in Figure~\ref{fig:non_bous_wave}($b$). This may be
seen to reduce the wave propagation speed since the speed is related to the
magnitude of the vorticity anomaly generation at the nodes\cite{Harnik-et-al08}.
The magnitude of the pressure anomalies and thus the resulting vorticity
anomalies are related by the size of $H$, which in this instance measures the
degree of deviation away from the Boussinesq limit. The same line of thought may
be applied to the $c<0$ case, which results in this $q\sim- p$, and the
corresponding scenario is illustrated in Figure~\ref{fig:non_bous_wave}($b$).
Again, the sign of the resulting vorticity anomaly is seen to be opposite to the
one given in Figure~\ref{fig:bous_wave}($b$). Thus, in this setting, the
physical picture is that the baroclinic torque associated with the
non-Boussinesq effects reduces the wave propagation speed via generation of
opposite signed vorticity anomalies to the ones generated by the Boussinesq term
at the nodes. This reduction is symmetric in magnitude for waves propagating in
either direction, since the pressure distribution associated with the waves with
$c>0$ and $c<0$ depends only on the distribution of $\zeta$ (as $p\sim -c\zeta$,
which may be seen when combining equations \ref{s3:system1} and \ref{s3:pq}). We
will see in the next section how the background shear affects the propagation of
right and left going waves in an asymmetric way.

%==============================================================================

\section{Edge wave dynamics}\label{s4}

Suppose now our background profiles are piecewise-continuous, so that
$\dy\Qbar/\dy z$ and $\mathrm{d}\rhobar / \mathrm{d}z$ (and so $N^2(z)$) are
defects of the form
\begin{equation}
	\ddy{\Qbar}{z}=\Delta\Qbar\delta(z-h),\qquad
	N^2(z)=\Delta N^2\delta(z-h).
\end{equation}
Then it may be seen that solutions of the form
\begin{equation}\label{s4:modal}
	q=\qhat\ex^{\zi k(x-ct)}\delta(z-h)
\end{equation}
are consistent solutions of \eqref{s2:system2} since there is no vorticity
generation away from the location of the defect at $z=h$. Taking also modal
solutions of $\psi$, $\zeta$ and $p$, we have
\begin{equation}\label{s4:equ1}
  (\Ubar-c)\qhat = -\Delta\Qbar\psi-\Delta N^2\zeta
  +\frac{\Delta N^2}{g}\frac{p}{\rhobar},\qquad
  (\Ubar-c)\zeta = \psi,
\end{equation}
where all the relevant terms are to be evaluated at $z=h$. In a domain that is
unbounded in $z$, $\psi$ is related to $q$ via a Green's function
\begin{equation}\label{s4:green}
  \psi(h)=-\frac{1}{2k}\int q(z')\ex^{-k|h-z'|}\, \mathrm{d}z'=
  -\frac{\qhat}{2k},
\end{equation}
and it remains to relate $p$ to the prognostic variables $q$ and $\zeta$.

We now wish to substitute $p/\rhobar$ for $u=-\dy\psi/\dy z$ by making use of
the $x$-momentum equation. Generically, $\psi$ is not differentiable at $z=h$,
however, $u$ changes sign when $z=h$ is crossed, hence, physically, for a wave
supported on $z=h$, there can be no self-induced $u$ for a wave-like solution,
and thus $u=0$ at $z=h$. With this, the $x$-momentum equation becomes in this
case
\begin{equation}\label{s4:qp-relation}
  0=-\ddy{}{x}\frac{p}{\rhobar}+\Qbar\ddy{\psi}{x}\qquad\Rightarrow\qquad
  \Qbar\psi=\frac{p}{\rhobar}.
\end{equation}
This means that, in the absence of shear, the pressure perturbation of the
interfacial wave is zero, and the Boussinesq approximation holds exactly in the
linearised baroclinc torque.

Substituting for $\psi$ and $p/\rhobar$ in \eqref{s4:equ1}, we obtain
\begin{equation}
	(\Ubar-c)\qhat=
	\left(\Delta\Qbar-\frac{\Qbar}{g}\Delta N^2\right)\frac{\qhat}{2k}
	-\Delta N^2\zeta, \qquad
	(\Ubar-c)\zeta=-\frac{\qhat}{2k}.
\end{equation}
The eigenstructure and dispersion relations are thus given by
\begin{equation}\label{s4:eigen}
	\qhat^{\pm}=2k(c^{\pm}-\Ubar)\zeta^{\pm},\qquad
	(c^{\pm}-\Ubar)=
	-\left(\frac{\Delta\Qbar-(\Qbar/g)\Delta N^2}{4k}\right)\pm
	\sqrt{\left(\frac{\Delta\Qbar-(\Qbar/g)\Delta N^2}{4k}\right)^{2}
	+\frac{\Delta N^2}{2k}}.
\end{equation}
The coefficient $(\Qbar/g)\Delta N^2$ measures the deviation away from the
Boussinesq limit. When $\Delta N^2=0$, we recover Rossby waves, while for
$\Qbar=\Delta \Qbar=0$, we recover the gravity waves in the Boussinesq
regime\cite{Harnik-et-al08}. The plus and minus branch are the branches where
the appropriate sign is taken.

\begin{figure}
\begin{center}
	\includegraphics[width=0.8\textwidth]{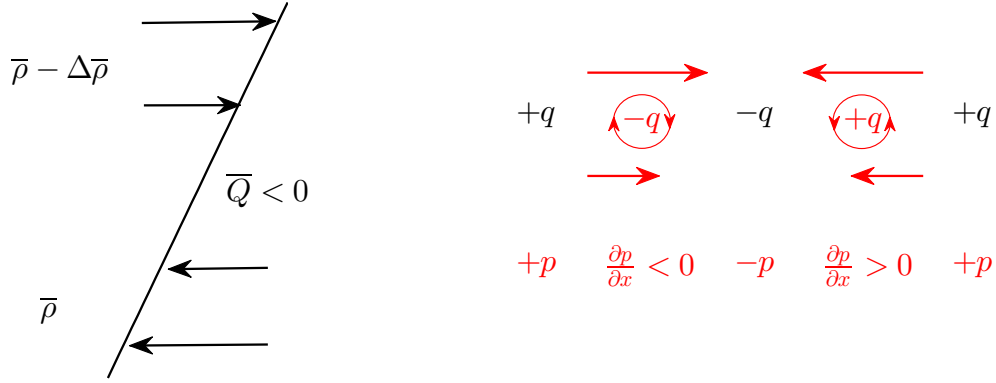}
	\caption{Schematic of the torque generated by the non-Boussinesq contribution
	for edge waves. With positive shear, we have $\Qbar<0$, and thus $-\psi\sim
	q\sim p$ from \eqref{s4:qp-relation}. The choice of shear fixes the relation
	between $q$ and $p$ at the peaks and troughs. With this, the associated
	pressure distribution generates vorticity anomalies at the nodes, introducing
	an asymmetry to the direction of wave propagation, in this case to leftward
	propagation.}
	\label{fig:non_bous_shear_wave}
\end{center}
\end{figure}

We make the observation that, for $(\Qbar/g)\Delta N^2\gg(\Delta
N^2,\Delta\Qbar)$, one of the branches vanish and so the edge waves become
uni-directional, with the preference direction dependent on the sign of $\Qbar$.
This is like the case for the propagation of Rossby waves, where in the absence
of stratification, is uni-directional and depends on the sign of $\Delta\Qbar$.
This preference for the direction of propagation is caused by the extra
contribution to the baroclinic torque from the non-Boussinesq term. To see how
this operates, we consider the case where we have a positive shear with
$\Qbar<0$. With this chioce, $q\sim-\psi\sim p$ from \eqref{s4:qp-relation} and
this fixes the $p$ and $q$ relation at the peaks and troughs, as displayed in
Figure~\ref{fig:non_bous_shear_wave}. Then we may consider both the case where
$q\sim\zeta$ (for rightward propagating waves) and $q\sim-\zeta$ (for leftward
propagating waves). Taking into account Figure~\ref{fig:bous_wave}($b$), it may
be seen that the resulting pressure anomalies results in vorticity anomalies at
the nodes that is of the opposite sign to the base case for the rightward
propagating wave (base case of Figure~\ref{fig:non_bous_wave}($a$)), i.e.,
counteracts propagation; on the other hand, the vorticity anomalies at the nodes
for the leftward propagation wave is the same sign as the base case (of
Figure~\ref{fig:non_bous_wave}($b$)), i.e., reinforcing propagation. Thus there
is a leftward preference for wave propagation due to the non-Boussinesq
contribution with positive shear; this is seen to be consistent with the
dispersion relation given in \eqref{s4:eigen}.

In this scenario with edge waves, the $p$ and $q$ relation is fixed by the sign
of the shear, where as in the Section~\ref{s2} for neutral waves supported in a
flow with no shear, the distribution depends on the direction of wave
propagation. The correction to the baroclinic torque acts to counteract wave
propagation in a symmetric way for the neutral wave case, whilst the presence of
a shear introduces a preferred direction for wave propagation.

%==============================================================================

\section{Non-Boussinesq Taylor--Caulfield instability}\label{s5}

One mechanistic interpretation for the onset of shear instabilities is via the
constructive interference of counter-propagating waves. Waves that propagate
vorticity anomalies may become phase-locked with each other via the advection by
the background flow and action-at-a-distance of the nonlocal velocity field
induced by local vorticity anomalies. With phase-locking, depending on the phase
shifts, these waves may amplify each other and lead to
instability\cite{Holmboe62, Harnik-et-al08, Rabinovich-et-al11,
Carpenter-et-al13}.

Since one of the key ingredients for this interpretation is counter-propagation,
our hypothesis with the previous section in mind is that, when the correction
term (as measured by with relation to $(\Qbar/g)\Delta N^2$) becomes
significant, instabilities reduce in growth rates and eventually switch off
because the waves can no longer phase-lock as they become increasingly
uni-directional. This suggests a physical interpretation to the work of Barros
\& Choi\cite{BarrosChoi14}, who find that a large shear across the interfaces
plays a stabilising role, which is perhaps somewhat counter-intuitive as the
shear is normally seen as a source of instability. To test this hypothesis, we
consider a simplified form of the the Taylor--Caulfield problem\cite{Taylor31,
Caulfield94, Rabinovich-et-al11, Balmforth-et-al12, GuhaLawrence14}, where the
basic state is essentially given by
\begin{equation}
  \Ubar(z)=\Lambda z, \qquad
  N^2(z) = \Delta N^2[\delta(z-h) + \delta(z+h)],
\end{equation}
with the $\delta$-functions in $N^2(z)$ coming from the choice that $\rhobar =
\rho_0 + |\delta\rhobar|[1-\mbox{H}(z-h)-\mbox{H}(z+h)]$, $\mbox{H}(z)$ the
Heaviside function, $h$ are the locations of the defects, and the imposed
density is a staircase-like profile. With this, the instability comes from the
interaction of two interfacial gravity waves located on the
defects\cite{Caulfield94, Rabinovich-et-al11, Balmforth-et-al12,
GuhaLawrence14}.

We proceed to non-dimensionalise the equations. By scaling with respect to
$T_0=\Lambda^{-1}$ and $L_0=h$, and taking modal solutions as in
\eqref{s4:modal}, it may be seen that the dimensional equations
\eqref{s2:system2} becomes (noting that $\delta$-functions have dimensions
$L_0^{-1}$ and that $\dy\Qbar/\dy z=0$ here)
\begin{equation}\label{s5:system-nondim}
	(\pm1 - c)\qhat_{1,2}	= -\Ri\left(\zeta_{1,2}
	-\Fr^2\left(\frac{p}{\rhobar}\right)_{1,2}\right),\qquad
	(\pm1 - c)\zeta_{1,2}=\psi_{1,2},
\end{equation}
where the equations are evaluated at $z=\pm1$ for subscript $1$ and $2$
respectively, and all quantities are non-dimensinonal. The non-dimensional
parameters in this case are
\begin{equation}
  \Ri = \frac{\Delta N^2}{h\Lambda^2},\qquad \Fr^2 = \frac{h^2\Lambda^2}{gh}.
\end{equation}
The Richardson number $\Ri$ measures the strength of the stratification. The
Froude number $\Fr$ is given by the square of the mean shear velocity scaled by
the Boussinesq gravity wave speed. Since the presence of shear allows the
non-Boussniesq baroclinic term to operate, it measures the deviation from the
Boussinesq limit. In the limit $\Fr\to0$, we recover the Boussinesq limit where
solutions to the problem as stated are known\cite{Rabinovich-et-al11}. With this
rescaling, the edge wave structure \eqref{s4:eigen} associated with this set up
is is given by
\begin{equation}\label{s5:eigen}
	\qhat^{\pm}_{1,2}=2k(c^{\pm}-\Ubar)_{1,2}\zeta^{\pm}_{1,2},\qquad
	(c^{\pm}-\Ubar)_{1,2}=
	-\frac{\gamma}{2}\pm \sqrt{\left(\frac{\gamma}{2}\right)^{2}
	+\frac{\Ri}{2k}}. \qquad\left(\gamma=\frac{\Ri\Fr^2}{2k}\right)
\end{equation}

It remains to relate $\psi_{1,2}$ and $(p/\rhobar)_{1,2}$ to $\qhat_{1,2}$ and
$\zeta_{1,2}$. First, $\psi_{1,2}$ may be related to $q_{1,2}$ via the Green's
function in an unbounded domain as in \eqref{s4:green}, except here we
have\cite{Harnik-et-al08, Rabinovich-et-al11}
\begin{equation}\label{s5:green}
  \psi_{1,2} = -\frac{1}{2k}(\qhat_{1,2} + \qhat_{2,1}\ex^{-2k}).
\end{equation}
Note that we have a term with a flipped subscript to denote the interaction
induced by anomalies on the other interface, with the exponential factor
representing the decay of interaction strength. For $(p/\rhobar)_{1,2}$, we
again make use of the $x$-momentum equation, which is, in this setting and with
$u=-\dy\psi/\dy z$,
\begin{equation}
  -(\pm1 - c)\ddy{\psi_{1,2}}{z} 
  = -\left(\frac{p}{\rhobar}\right)_{1,2}-\psi_{1,2}.
\end{equation}
The physical argument here is that there should be no self-induced $u$ on an
interface but there may be an induced $u$ from the other interface. Since a
positive vorticity anomaly induces a positive $u$ below and negative $u$ above
it (and vice-versa for negative vorticity anomalies), we obtain
\begin{equation}
  -\ddy{\psi}{z}=\begin{cases}
  -\frac{1}{2}\int q(z')\ex^{-k(z-z')}\, \mathrm{d}z', & z>z',\\
  +\frac{1}{2}\int q(z')\ex^{+k(z-z')}\, \mathrm{d}z', & z<z',
  \end{cases} \qquad\Rightarrow\qquad
  -\ddy{\psi_{1,2}}{z} = \mp\frac{1}{2}\qhat_{2,1}\ex^{-2k}.
\end{equation}
Substituting the above into \eqref{s5:system-nondim}, we obtain the governing
system of equations
\begin{equation}\label{s5:system-nondim1}
	(\pm1 - c)(\qhat_{1,2} \mp \gamma k \qhat_{2,1}\ex^{-2k}) = 
	-\Ri\zeta_{1,2} + \gamma(\qhat_{1,2} + \qhat_{2,1}\ex^{-2k}),\qquad
	(\pm1 - c)\zeta_{1,2} = -\frac{1}{2k}(\qhat_{1,2} + \qhat_{2,1}\ex^{-2k}).
\end{equation}

In matrix form, this is
\begin{equation}\label{s5:matrix}
	\begin{pmatrix}
	1-c-\gamma & \Ri & -\gamma\ex^{-2k}[1+(1-c)k] & 0\\
	1/(2k) & (1-c) & \ex^{-2k}/(2k) & 0\\
	-\gamma\ex^{-2k}[1+(1+c)k] & 0 & -(1+c+\gamma) & \Ri\\
	\ex^{-2k}/(2k) & 0 & 1/(2k) & -(1+c)
	\end{pmatrix}\begin{pmatrix}
	\qhat_{1} \\ \zeta_{1} \\ \qhat_{2} \\ \zeta_{2}\end{pmatrix} = 0,
\end{equation}
and this yields the dispersion relation
\begin{equation}\label{s5:dispersion}\begin{aligned}
	&(1+\gamma^{2}\ex^{-4k}k^{2})c^{4}+2\gamma c^{3}
	-\left[2+(\Ri/k)
	-\gamma^{2}(1-\ex^{-4k}(1+2k+2k^{2}))\right]c^{2}\\
	&\qquad -\left[2\gamma(1-\Ri\ex^{-4k}) + \gamma(\Ri/k)(1-\ex^{-4k})\right]c\\
	&\qquad\qquad +\left[1-\frac{\Ri}{k} +\frac{\Ri^{2}}{4k^{2}}(1-\ex^{-4k})
	+\gamma^{2}(\ex^{-4k}(1+k)^{2}-1)\right]=0.
\end{aligned}\end{equation}
When $\Fr=0$, $\gamma=0$, and \eqref{s5:matrix} as well as \eqref{s5:dispersion}
reduce to previously known forms in the Boussinesq
limit\cite{Rabinovich-et-al11}.

\begin{figure}[tb]
\begin{center}
	\includegraphics[width=0.8\textwidth]{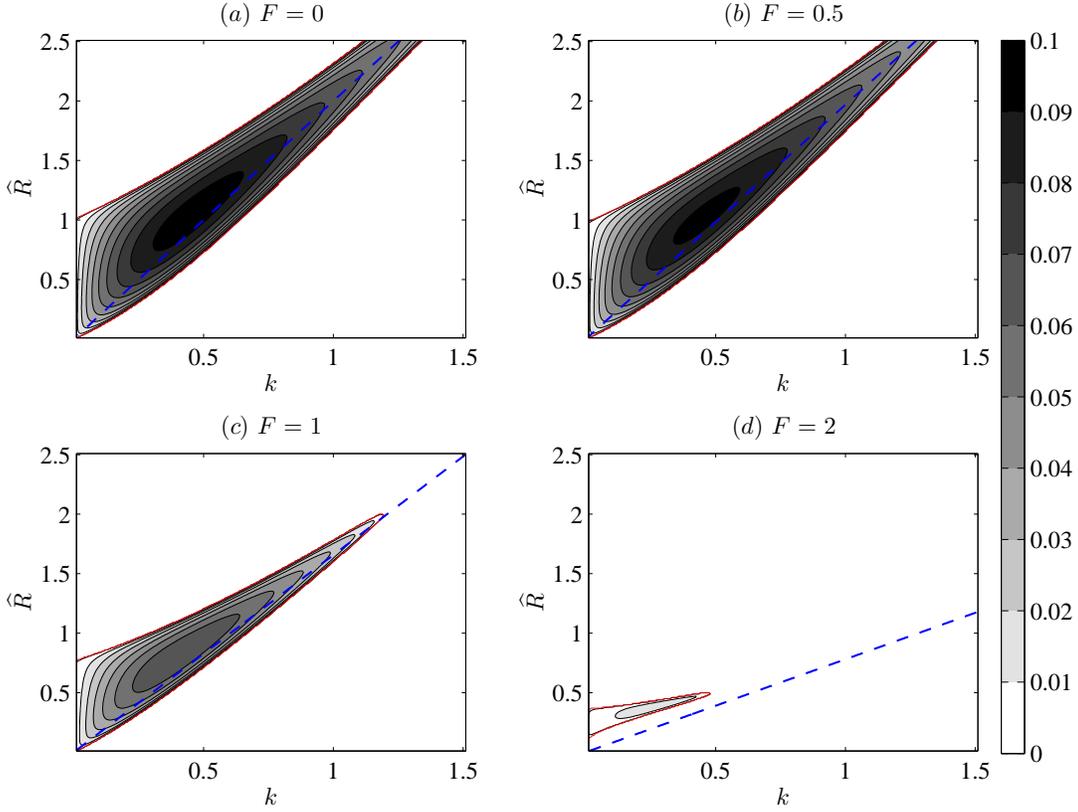}
	\caption{Contour of growth rate $kc_i$ in ($k,\Ri$) space at various values of
	$\Fr$; all the contours are at fixed levels at $0.01$ spacing for all four
	panels. The blue dashed contours are where the resonance condition
	\eqref{s5:resonance} is satisfied.}
	\label{fig:TC_max_growth}
\end{center}
\end{figure}

The dispersion relation \eqref{s5:dispersion} may be solved numerically to
obtain the four roots and this was done using the MATLAB command \verb!roots!.
We first show in Figure~\ref{fig:TC_max_growth} the contours of the growth rates
over $(k,\Ri)$ space at several values of $\Fr$.
Figure~\ref{fig:TC_max_growth}($a$) is exactly the solution in the Boussinesq
limit for which an analytic expression for the solution is
available\cite{Rabinovich-et-al11}. As we increase $F$, the growth rates reduces
in Figure~\ref{fig:TC_max_growth}($b$), notably around the region of maximum
growth. As $\Fr$ is increased further, the maximum growth rate decreases, and
the region of instability shrinks towards the small $k$ region, as seen in
Figure~\ref{fig:TC_max_growth}($c,d$).

Sometimes it is useful to show the locations where the resonance condition is
satisfied\cite{Carpenter-et-al13}. These are the locations where the
counter-propagating edge waves have matching phase speeds, taking into account
advection by the background flow. From equation \eqref{s5:eigen},
these are the values of $k$ where
\begin{equation}\label{s5:resonance}
  c_1^-(\Ri,\Fr,k) = 1 + 
  \left(-\frac{\gamma}{2} - \sqrt{\left(\frac{\gamma}{2}\right)^{2}
	+\frac{\Ri}{2k}}\right) \qquad \textnormal{and} \qquad
	c_2^+(\Ri,\Fr,k) = -1 + 
	\left(-\frac{\gamma}{2} + \sqrt{\left(\frac{\gamma}{2}\right)^{2}
	+\frac{\Ri}{2k}}\right)
\end{equation}
are equal. If interacting counter-propagating edge waves contribute the most to
the dynamics, then the location where the resonance condition is satified should
be near to the location of optimal growth; otherwise, it shows that other
dynamics (e.g. pro-propagating modes, critical layers) are important. It also
gives an indication of where in parameter space the interaction required for
instability may be expected. Locations of these are shown as dashed contours in
Figure~\ref{fig:TC_max_growth}, and we see these show reasonable correlation to
the locations of largest growth. However, we notice that, in the larger $\Fr$
cases, even though we may have edge waves with matching phase speeds, this does
not necessary indicate instability, since the resonance condition does not take
into account the mutual interaction.

\begin{figure}[tb]
\begin{center}
	\includegraphics[width=0.9\textwidth]{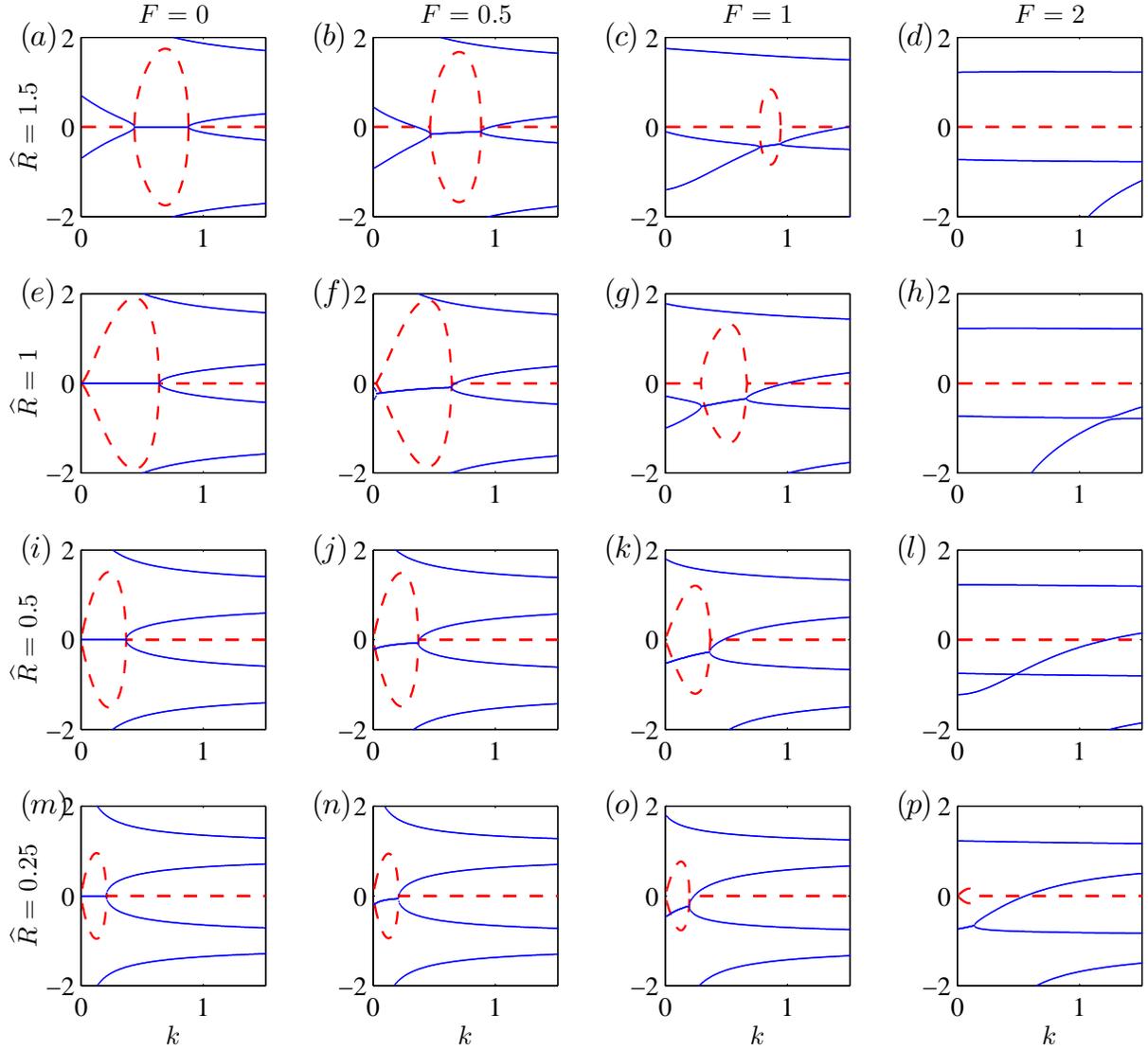}
	\caption{Line graphs of the solution branches of the dispersion relation
	\eqref{s5:dispersion}, with blue solid lines denoting $c_r$, and red dashed
	line denoting $20kc_i$ (factor of $20$ to emphasise the instability region),
	at different values of $\Ri$ and $\Fr$. The rows are at different values of
	$\Ri$ whilst the columns are at different values of $\Fr$ (see diagram).}
	\label{fig:TC_line}
\end{center}
\end{figure}

It is perhaps informative to see the behaviour of the individual solution
branches. In Figure~\ref{fig:TC_line} we show the phase speeds $c_r$ (solid
blue) and the (magnified) growth rate $20kc_i$ (dashed red) for several values
of $\Ri$ and $\Fr$. The first column shown in panels ($a,e,i,m$) is the
Boussinesq case where there is no preference for the direction of wave
propagation for this choice of basic state. Focusing on the phase speed, the
outer two branches are the neutral waves, while the inner branches represent the
stable and unstable branch, occurring in conjugate pairs. As $F$ increases, the
degree of asymmetry increases, with a preference for leftward propagating waves,
which is consistent with the result in Section~\ref{s4}. We also note that
although the branches appear to cross, there is only instability when a
pitchfork-like (rather than a transcritical one as in panels ($d,h,l$))
bifurcation occurs.

To further quantify the asymmetry between the leftward and rightward propagating
waves, we wish to obtain the instability in terms of the left and rightward
propagating modes $\zeta_{1,2}^\pm$. Unlike the previous formulations where
transformation matrices were present\cite{Harnik-et-al08, Rabinovich-et-al11},
the complication here is from the $(p/\rhobar)_{1,2}$ term in equation
\eqref{s5:system-nondim}. This contributes a $(\dy/\dy t+\Ubar\dy/\dy x)u$ term,
which means we can no longer write the problem in the form
$\dy\boldsymbol{\zeta}/\dy t = \boldsymbol{\mathsf{A}}\boldsymbol{\zeta}$ in a
simple way, and the transformation matrix acting on $\boldsymbol{\mathsf{A}}$
becomes complicated. In principle, since everything is linear, an alternative
approach that one could take is to work out how the individual terms in the
governing equation \eqref{s5:system-nondim} should look like, and the equations
for $\zeta_{1,2}^\pm$ should have on the right hand side the interaction terms
written in terms of the appropriate contributions from the terms in
\eqref{s5:system-nondim}. We may postulate for example that the equation for
$\zeta_1^+$ say should only be affected by all variables not including
$\zeta_1^-$, i.e., the governing equations without the modal solution assumption
should be of the form
\begin{equation}
	(\pm1-c)\qhat_{1,2}^\pm
	=-\Ri_{1,2}\left[\zeta_{1,2}^\pm
	-\Fr^2\left(\frac{p}{\rhobar}\right)^*\right],\qquad 
	(\pm1-c)\zeta_{1,2}^\pm	=\ddy{\psi^*}{x},
\end{equation}
where
\begin{equation}
  \psi^*=\psi_{1,2}^\pm+\left(\psi_{2,1}^+ + \psi_{2,1}^-\right)\ex^{-2k},
\end{equation}
and $(p/\rhobar)^*$ is to be defined analogously. With this, we may substitute
accordingly noting that: (i) $\qhat_{1,2}^\pm \rightarrow
2k(c^{\pm}-\Ubar)_{1,2}\zeta_{1,2}^{\pm}$ via the eigenstructure
\eqref{s5:eigen}; (ii) we take $\psi_{1,2}^\pm=-\qhat_{1,2}^\pm/2k$, and again
may be written in terms of $\zeta_{1,2}^{\pm}$ via the eigenstructure; (iii)
$\zeta_{1,2}^\pm$ is a local variable so we leave it as is; (iv) some care needs
to be taken for the $(p/\rhobar)^*$ term, but we essentially use the definition
that
\begin{equation}
	\frac{p(z)}{\rhobar}=-\psi-\begin{cases}0, &z=z_b\\
	(c(z_b)-\Ubar(z))(\dy\psi/\dy z), &z\neq z_b,\end{cases}
\end{equation}
and these may be then be written in terms of $\zeta_{1,2}^{\pm}$ via appropriate
substitutions.

The resulting manipulations are quite unwieldy due to the large number of terms
and we shall not present them here. Instead, we may achieve the same goal by
decomposing the resulting unstable modes into its normal modes. Since we already
have $c$ from the calculations, one way to do this is to write
\eqref{s5:matrix} as
\begin{equation}\label{s5:matrix-normal}
	\begin{pmatrix}
	1-\gamma & \Ri & -\gamma\ex^{-2k}[1+(1-c)k] & 0\\
	1/(2k) & 1 & \ex^{-2k}/(2k) & 0\\
	-\gamma\ex^{-2k}[1+(1+c)k] & 0 & -(1+\gamma) & \Ri\\
	\ex^{-2k}/(2k) & 0 & 1/(2k) & -1
	\end{pmatrix}\begin{pmatrix}
	\qhat_{1} \\ \zeta_{1} \\ \qhat_{2} \\ \zeta_{2}\end{pmatrix} = 
	\tilde{c}\begin{pmatrix}
	\qhat_{1} \\ \zeta_{1} \\ \qhat_{2} \\ \zeta_{2}\end{pmatrix},
\end{equation}
and solve for $\tilde{c}$ and $(\qhat_{1}, \zeta_{1}, \qhat_{2}, \zeta_{2})$
using the MATLAB \verb!eig! command, but returning a solution only
$|\tilde{c}-c|<10^{-15}$ (this condition is satisfied for all unstable solutions
computed here). To then transform the resulting solution into $\zeta_{1,2}^\pm$,
we make use of the wave structure \eqref{s5:eigen}, so that the unstable mode in
terms of normal modes is given by\cite{Harnik-et-al08}
\begin{equation}\label{s5:normal-decomp}
  \begin{pmatrix}\zeta_1^+ \\ \zeta_1^- \\ \zeta_2^+ \\ \zeta_2^-\end{pmatrix}
  = \begin{pmatrix}
	2k(c^+ - \Ubar)_1 & 2k(c^- - \Ubar)_1 & 0 & 0\\
	1 & 1 & 0 & 0\\
	0 & 0 & 2k(c^+ - \Ubar)_2 & 2k(c^- - \Ubar)_2\\
	0 & 0 & 1 & 1
	\end{pmatrix}^{-1}
	\begin{pmatrix}\qhat_{1} \\ \zeta_{1} \\ \qhat_{2} \\ \zeta_{2}\end{pmatrix}.
\end{equation}
The plus and minus superscripts denote the rightward and leftward
propagating modes, and it is primarily the interaction between the two
counter-propagating modes $\zeta_1^-$ and $\zeta_2^+$ that leads to instability,
with the pro-propagating modes $\zeta_1^+$ and $\zeta_2^-$ that modify the
interactions accordingly.

\begin{figure}[tb]
\begin{center}
	\includegraphics[width=0.9\textwidth]{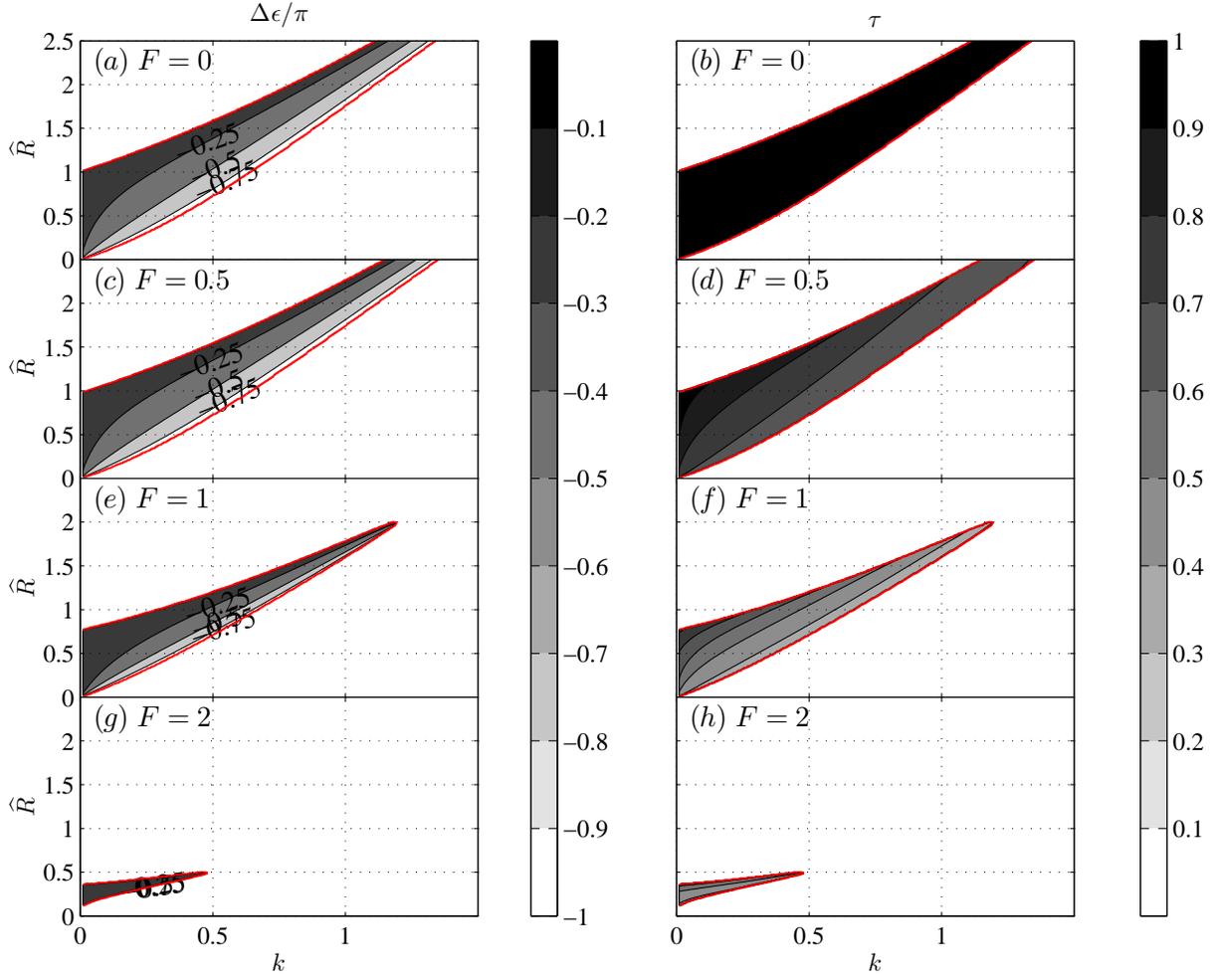}
	\caption{Diagnostics from the normal modes. ($a,c,e,g$) shows the normalised
	phase difference $\Delta\epsilon/\pi$, where $-0.5<\Delta\epsilon/\pi<0$ is
	the `unstable hindering' regime, and $-1<\Delta\epsilon/\pi<-0.5$ is the
	`unstable helping' regime. ($b,d,f,h$) shows the ratio of the total amplitude
	of rightward-propagating modes and the total amplitude of leftward-propagating
	modes.}
	\label{fig:TC_diagnostics}
\end{center}
\end{figure}

With
\begin{equation}\label{s5:amp-phase}
  \zeta_{1,2}^\pm = A_i^\pm \ex^{\zi\epsilon_{1,2}^\pm},
\end{equation}
we define $A_i^\pm>0$ and $\epsilon_{1,2}^\pm\in(-\pi,\pi]$ to be the (real)
amplitude and phase of the respective modes. With this, we show in
Figure~\ref{fig:TC_diagnostics} the (normalised) phase difference
$\Delta\epsilon/\pi$ between the counter-propagating modes and the ratio of the
amplitude of the rightward-propagating waves and the leftward-propagating waves
$\tau$, respectively given by
\begin{equation}\label{s5:diagnostics}
  \Delta\epsilon = \epsilon_2^+ - \epsilon_1^-,\qquad
  \tau = \frac{A_1^+ + A_2^+}{A_1^- + A_2^-}.
\end{equation}

\begin{figure}[tb]
\begin{center}
	\includegraphics[width=\textwidth]{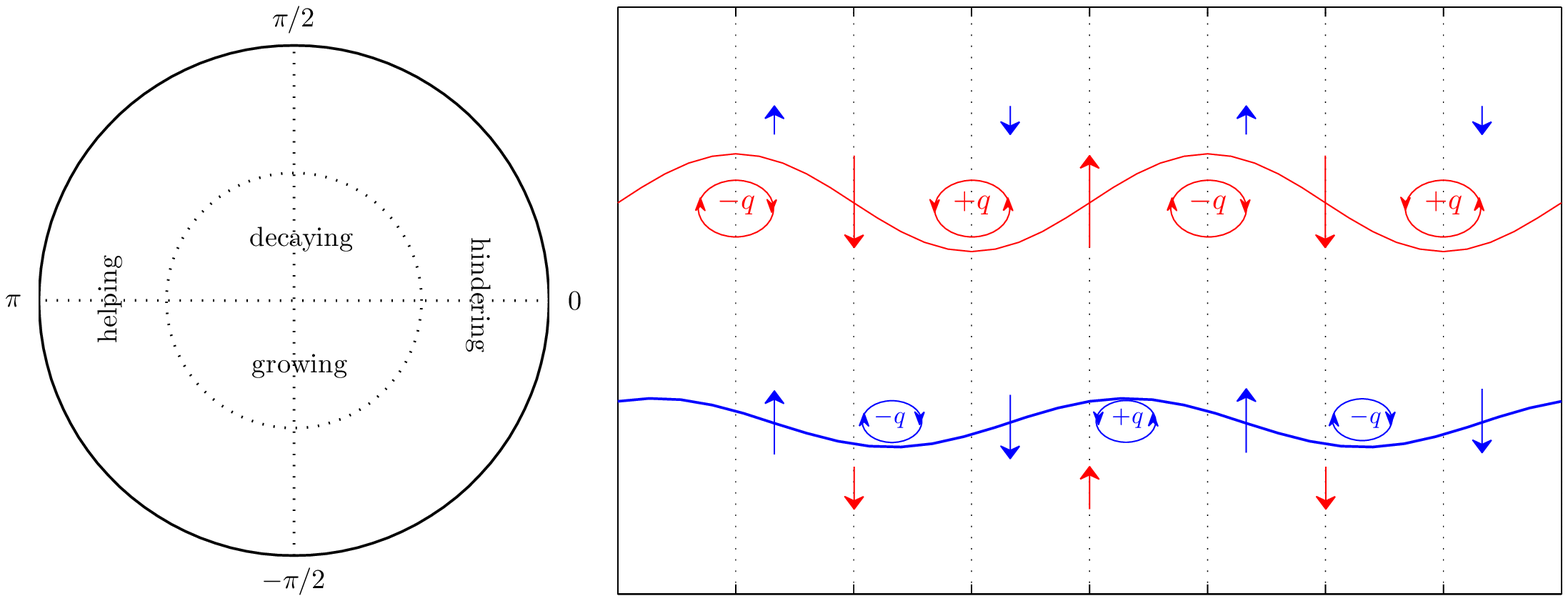}
	\caption{A regime diagram and for phase differences and the likely physical
	scenario of the resulting instability in the non-Boussinesq regime. The regime
	diagram is defined using displacement with phase difference as defined in
	equation \eqref{s5:diagnostics}. The wave configuration between the two
	counter-propagating modes is in a growing hindering regime here, with the top
	wave travelling to the left and bottom wave travelling to the right. The top
	wave is of large amplitude, as indicated by the larger labels and arrows,
	consistent with the diagnostics as in Figure~\ref{fig:TC_diagnostics}.
	Although the waves are phase-locked, the leftward propagating wave dominates
	and imparts a leftward propagation to the resulting instability, consistent
	with the observations in Figure~\ref{fig:TC_line}.}
	\label{fig:CRW_config}
\end{center}
\end{figure}

Starting first with the phase difference, since we defined it using the
displacement rather than vorticity\cite{Heifetz-et-al15}, it may be seen that
$-\pi<\Delta\epsilon<0$ is the unstable regime, and with
$-\pi/2<\Delta\epsilon<0$, we are in the `hindering' regime where the
configuration is such that the counter-propagating waves hinder each other's
propagation to achieve phase-locking, typical of fast waves (see also
Figure~\ref{fig:CRW_config} here). We see in panels ($a,d,g,j$) this occurs for
waves at higher values of $\Ri$ and small $k$, which is consistent with the
dispersion relation \eqref{s5:eigen}, where faster waves occurs for larger $\Ri$
and smaller $k$. The reverse is true when $-\pi<\Delta\epsilon<-\pi/2$, and we
are in the `helping' regime. This explains why the locations where the resonance
condition \eqref{s5:resonance} is satisfied does not necessarily correspond to
the location of largest growth rate. As indicated from the Green's function
\eqref{s4:green} and \eqref{s5:green}, the interaction strength between the
waves increases as $k$ decreases, but then so does the counter-propagation speed
from \eqref{s4:eigen} and \eqref{s5:eigen}. Hence, the gravest mode is obtained
in growing, hindering configurations, which is a generic result that applies to
barotropic and baroclinic instabilities\cite{Heifetz-et-al99,
Heifetz-et-al04a}.

For the ratio $\tau$ as defined in \eqref{s5:diagnostics}, we make the
observation that, for the Boussinesq limit where $\Fr=0$, there is no preference
for direction of wave-propagation, so the value of $\tau$ should be equal to $1$
over the parameter space, which is what we see in panel ($b$). For non-zero
$\Fr$, there is a preference for leftward-propagation, so
the value of $\tau$ is less than $1$ and decreases in size as $\Fr$ increases,
which is what we observe in panels ($d,f,h$). We make the observation that the
asymmetry is less strong for long-waves, indicating the non-Boussinesq effect
appears to have a stronger effect on short-waves. This is perhaps consistent
with the expectation that we expect buoyancy effects to remain dominant for
large-scale motions, and non-Boussinesq term affect small-scale motion more
substantially. Notice that the $\tau$ does not need to vanish for the
instability to switch off; waves being unidirectional is a sufficient but not
necessary condition for phase-locking, and the ability to phase-lock may
disappear before waves become unidirectional.

In the work of \citeauthor{Rabinovich-et-al11}\cite{Rabinovich-et-al11} in the
Boussinesq regime, it was argued that, for phase-locking, the pro-propagating
mode on one flank should be in anti-phase and smaller by a factor of $\chi$ with
the counter-propagating mode on the other flank. The picture is likely to be
somewhat more complicated here in the non-Boussinesq regime. There is now a
preference for the direction of travel, and thus $\chi_1 = A_2^-/A_1^-$ may not
be (and is generically not) equal to $\chi_2 = A_1^+/A_2^+$. These diagnostics
do not tell us anything overly meaningful, and a presentation of the associated
results has been omitted. For completeness, the ratio between the two
counter-propagating modes $\chi_0 = A_2^+/A_1^-$ shows that they are mostly
comparable in magnitude over the unstable region, with a slight preference
towards the leftward counter-propagating mode $A_1^-$. As we have seen before,
the observed instabilities are no longer stationary modes, and thus a clear
picture as in \citeauthor{Rabinovich-et-al11} is unlikely to hold in this case.
A likely physical scenario for instability is that phase-locking is still
achieved, but, as we may expect from the hypothesis, since there is a preference
for leftward propagation, the leftward travelling counter-propagating mode
dominates and imparts a leftward propagation to the resulting instability, which
is consistent with the negative values of $c_r$ observed in
Figure~\ref{fig:TC_max_growth}. A schematic of the resulting interaction between
counter-propagating waves is shown in Figure~\ref{fig:CRW_config}, and we expect
this to be the fundamental component in driving the instability, with the
pro-propagating modes modifying the interaction in a more complicated manner.

%==============================================================================

\section{Conclusion and discussion}\label{s6}

In this article, we investigated how the portion of the baroclinic torque that
is neglected by the Boussinesq approximation affects wave propagation, and how
this in turn affects the mechanistic interpretation for shear instability.
Increasingly complex examples were considered and rationalised, and it was found
that the dynamics depends on the relation between pressure and vorticity
anomalies. In Section~\ref{s3} we observed that, for neutral waves supported on
a uniform background flow, the non-Boussinesq term acts to reduce the wave speed
via generating vorticity anomalies that counteract what would otherwise be
generated by the Boussinesq term, in a symmetric way that depends only on the
direction of wave propagation. In Section~\ref{s4}, the introduction of a
background shear fixes this degree of freedom between the pressure and vorticity
anomalies and introduces an asymmetry for direction of wave propagation, which
is to the left for positive shear ($\Qbar<0$). In Section~\ref{s5}, a simplified
version of the Taylor--Caulfield problem was investigated and analysed. With
positive shear, the hypothesis was that, since there is a preference for
leftward propagation of waves, as we increase the non-Boussinesq effect as
measured by the Froude number $\Fr$, the waves should become increasingly
uni-directional. With this, phase-locking becomes harder to achieve, and thus
increasing $\Fr$ reduces the region of instability and the maximum growth rates.
This was indeed found to be the case via plots of the maximum growth rate in
Figure~\ref{fig:TC_max_growth} and the values of the ratio of the total
rightward propagating waves to the total leftward propagating waves $\tau$ shown
in Figure~\ref{fig:TC_diagnostics}, the latter obtained by a decomposition of
the unstable modes into its left and rightward propagating constituents via the
dispersion relation \eqref{s5:eigen}. 

These results are in general agreement with the previous works on shear
instability in non-Boussinesq systems\cite{UmurhanHeifetz07, BarrosChoi11,
BarrosChoi14} even if their precise set up is not identical to ours. In
particular, the observation that increasing the shear (i.e., the value of $\Fr$)
stabilises the instability\cite{BarrosChoi14} is in agreement of our results
here, with the reason being that the magnitude of the shear increases the degree
of asymmetry for wave propagation, which in turn affects phase-locking
properties. Furthermore, the work of Barros \& Choi\cite{BarrosChoi14} analysed
the non-Boussinesq effect when it is combined with the effect of confinement by
boundaries. As shown in some previous works\cite{Heifetz-et-al09,
BiancofioreGallaire12}, the reduction in growth owing to confinement can also be
explained in terms of wave interaction, since mirror image waves that are in
anti-phase with the counter-propagating waves may be placed on the other side of
the boundaries to enforce the boundary conditions accordingly, as in the method
of images. This results in a reduction of the overall interaction strength as
well as the ability of each wave to counter-propagating against the mean flow.

We believe that our observations and interpretation carries over to the
non-Boussinesq Holmboe\cite{Holmboe62} problem investigated
previously\cite{UmurhanHeifetz07, BarrosChoi11, BarrosChoi14}, which is often
attributed to the interaction between a Rossby wave and a gravity
wave\cite{Carpenter-et-al10a, Balmforth-et-al12, Carpenter-et-al13,
GuhaLawrence14}, with wave speed governed by $\Delta\Qbar$ and $\Ri$
respectively. In the piecewise-linear set up as in Holmboe's original set up,
the non-Boussinesq term affects the gravity waves supported on the density
defect within the shear layer but not the Rossby waves, since $\Delta N^2=0$ at
the location of the vorticity defects. Normally there is a symmetric Holmboe
mode arising from the interaction between a leftward Rossby wave with a
rightward gravity wave, together with a rightward Rossby wave and a leftward
gravity wave. Non-Boussinesq effects will modify the gravity waves so that the
interaction is no longer symmetric, and the instabilities should have non-zero
$c_r$, as in some previous works where the interaction was made asymmetric via
other means (e.g., making the distance between the density and the vorticity
defects asymmetric)\cite{HaighLawrence99, Alexakis05, Alexakis07,
UmurhanHeifetz07, Carpenter-et-al07, Tedford-et-al09, Carpenter-et-al10a,
BarrosChoi11, GuhaLawrence14, BarrosChoi14}. We expect an analogous schematic to
the one shown in Figure~\ref{fig:CRW_config} should hold for the Holmboe
problem. Similar effects should also be observed when smooth basic
states\cite{HaighLawrence99, Alexakis05, Alexakis07, Carpenter-et-al07,
Alexakis09, Tedford-et-al09} are considered in the non-Boussinesq regime. In
terms of general applicability, since large-scale stratified flows tend to be
dominated by buoyancy effects, non-Boussinesq effects are more likely to
manifest for small-scale flows. Since such instabilities has been observed to
lead to mixing\cite{SmythWinters03, Smyth-et-al07, Carpenter-et-al07} the
non-Boussinesq effects on these instabilities may indirectly affect the mixing
properties, although this possible avenue for further research is beyond the
scope of this present study.

%==============================================================================

%\label{}
%\subsection{}
%\subsubsection{}

% If in two-column mode, this environment will change to single-column format so that long equations can be displayed. 
%\begin{widetext}
%$$\mbox{put long equation here}$$
%\end{widetext}

% Figures should be put into the text as floats. 
% Use the graphics or graphicx packages (distributed with LaTeX2e).

% Put the label that you will use with \ref{} command in the braces of the \label{} command.
%
% \begin{figure}
% \end{figure}


% \begin{table}
% \end{tabular}
% \end{table}


\begin{acknowledgments}

% Put your acknowledgments here.

JM was supported by the Israeli Science Foundation grant 1537/12 and the UK NERC
grant NE/L005166/1 for the duration of this work. We thank Abigail Bodner, Nili
Harnik and Ron Yellin for discussions relating to this work. The authorship is
alphabetical.

\end{acknowledgments}

%=====================================================================================
%=====================================================================================


%\bibliographystyle{unsrtnat}
%\bibliographystyle{plainnat}
%\bibliography{paper4}

\bibliography{refs}

\end{document}